\documentclass[10pt,conference,table]{IEEEtran}
\usepackage{amsmath}
\usepackage{amssymb}
\usepackage{graphicx}
\usepackage{tikz}
\usepackage{booktabs}
\usepackage[noadjust]{cite}

\usetikzlibrary{arrows.meta, positioning, shapes.geometric, calc}
\usepackage{xspace}
\usepackage{needspace}

\usepackage{paralist}

\usepackage[absolute,overlay]{textpos} 

\newcommand{\mycheck}{\cellcolor{green!50}\checkmark}

\newcommand{\method}{\textsc{NetCause}\xspace}
\newcommand{\hide}[1]{}
\newcommand{\bulletEmph}[1]{{\bf #1}}
\newcommand{\shortTag}[1]{{\bf #1}}

\newcommand{\effectiveness}{\bulletEmph{Effectiveness}\xspace}

\newcommand{\metric}{\bulletEmph{Metric}\xspace}
\newcommand{\model}{\bulletEmph{Model}\xspace}
\newcommand{\formulation}{\bulletEmph{Problem Formulation}\xspace}

\begin{document}
\IEEEoverridecommandlockouts
\IEEEpubid{\hspace{-15.5cm}$^{*}$Equal contributions.}

\title{\method: Counterfactual Learning for Root Cause Analysis in Large-Scale Networks}

\author{\IEEEauthorblockN{Fabien Chraim\textsuperscript{*}, Jian Zhang\textsuperscript{*}, Dominik Janzing, Xiang Song, Christos Faloutsos, John Evans}
\IEEEauthorblockA{Amazon Web Services, Seattle, Washington, USA\\
Email: chraim@amazon.com, jamezhan@amazon.com, janzind@amazon.de, \\
xiangsx@amazon.com, faloutso@amazon.com, jevanamz@amazon.co.uk}}

\maketitle

\begin{textblock*}{0.9\textwidth}(0.12\textwidth,1.1\textheight)
\centering
\footnotesize
© 2026 IEEE. This is the author's preprint version of a paper accepted for publication in the Proceedings of IEEE International Conference on Computer Communications and Networks, 2026. The final version will appear in IEEE Xplore.
\end{textblock*}

\begin{abstract}
Can a learned model capture how faults propagate through a large-scale network and use this knowledge to causally attribute customer impact to its underlying root cause? Existing root cause analysis techniques often rely on static rules, correlation heuristics, or topology-local reasoning, which struggle to generalize in dynamic environments where faults propagate across complex physical and logical dependencies.

We present \method, a self-supervised learning-based framework that models network incidents as graph-temporal processes and uses counterfactual simulation to rank candidate root causes. This approach produces an interpretable ranking of root cause hypotheses and integrates naturally with operator-defined mitigation and remediation actions.

We train the model on over 1{,}500 incidents collected over six months from a leading cloud provider's production network and evaluate it on 31 expert-labeled incidents. \method consistently improves root cause ranking quality in the regime most relevant to operational decision-making, achieving a \underline{16.1\%} accuracy improvement over a rule-based heuristic baseline. While training is computationally intensive, inference is lightweight, requiring only \underline{seconds} of GPU runtime per incident (well below typical telemetry collection latencies).

\end{abstract}


\section{Introduction}

\begin{figure}[!t]
\centering
\includegraphics[width=\columnwidth]{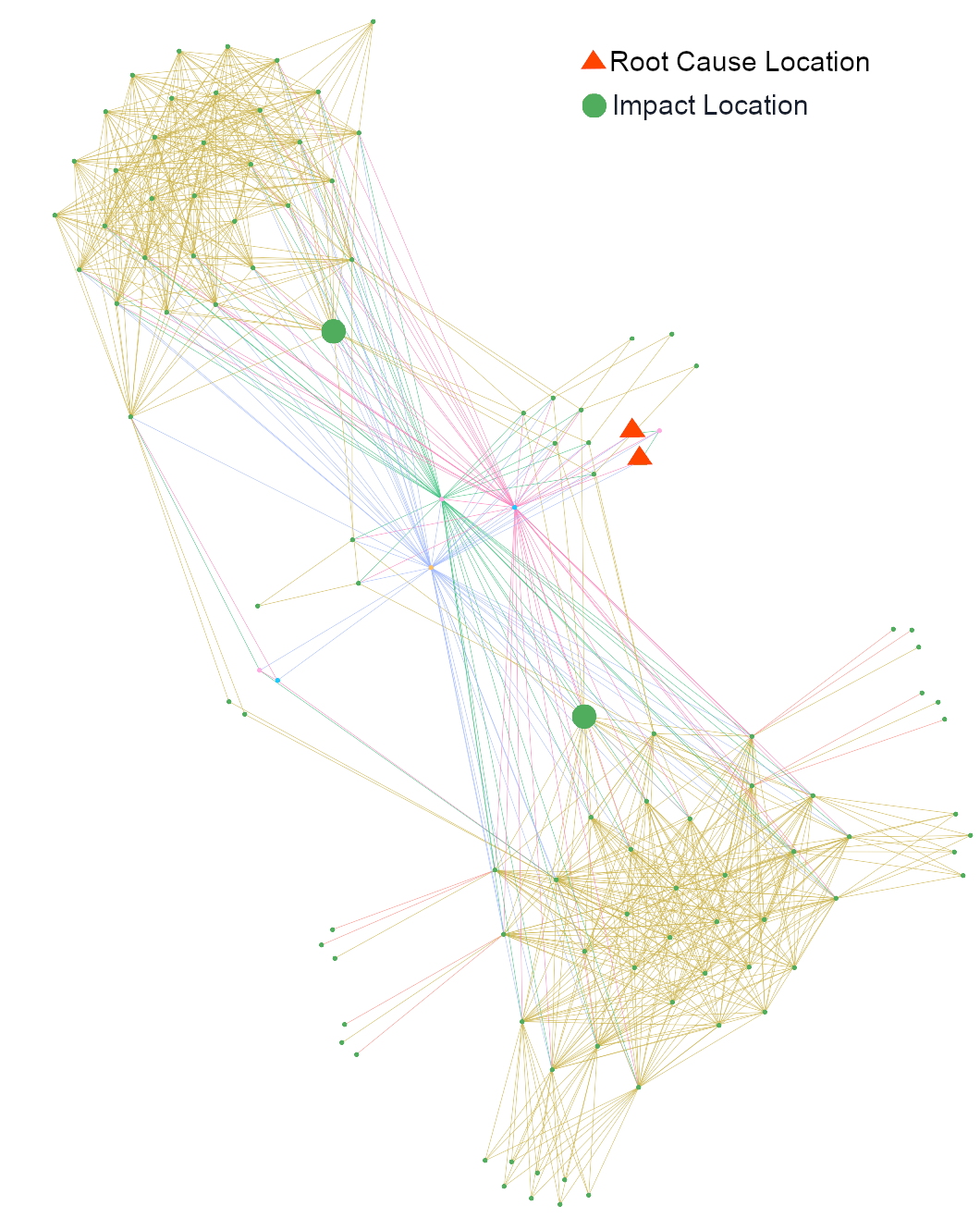}
\caption{\textit{Example incident subgraph from a production cloud network}. Nodes represent network devices and hierarchical aggregations, and edges denote physical and logical relationships (e.g. yellow edges represent layer 3 neighborship; blue edges show metro containment). Customer impact is observed on the two green nodes, corresponding to routers. The underlying root cause is a reachability failure on the red triangular nodes, which correspond to controller devices. These events occur minutes apart and are not topological neighbors; they are separated by two hops through a hierarchical aggregation node. This example illustrates the challenge of RCA in large-scale cloud networks, where causally related signals may be temporally and topologically distant within complex, heterogeneous topologies.}
\label{fig:incident-graph}
\end{figure}

Modern cloud infrastructure operates at unprecedented scale, with millions of interconnected devices spanning data centers, metro networks, and wide-area backbones. When customer-impacting incidents occur, operators must identify root causes amid cascading fault propagation across complex physical and logical topologies in order to remediate underlying failures. Traditional root cause analysis (RCA) approaches rely on manually crafted correlation rules or proximity-based heuristics. These struggle to capture the intricate spatiotemporal dependencies inherent in large-scale network incidents. A central challenge is distinguishing causally related faults from those that merely co-occur in spatial or temporal proximity. This issue is compounded by the sheer volume of telemetry signals and noise generated by modern infrastructure.

A network incident typically manifests as a sequence of observable events, such as device faults, link failures, and traffic migrations, that propagate through the system before culminating in customer-visible impact (e.g., packet loss or latency). While temporal precedence and topological proximity provide useful cues, they are insufficient for causal attribution. A fault that occurs early and near the impact may be symptomatic rather than causal, while the true root cause may be temporally or spatially distant. Moreover, multiple independent faults often coexist within large networks, further complicating attribution.

To illustrate these challenges, Figure~\ref{fig:incident-graph} depicts an incident subgraph extracted from a production cloud network. Nodes represent network devices and hierarchical aggregations, while edges capture physical and logical relationships. For example, yellow edges shot OSI layer 3 neighborship while blue edges represent the membership of a node in a metro. The subgraph is constructed by expanding a two-hop neighborhood around impact nodes where packet loss was detected (shown as the two larger green nodes). Although the resulting subgraph contains 113 nodes and 868 edges, the customer impact is ultimately caused by a reachability failure affecting only two nodes (highlighted as red triangles).

For clarity, we omit additional observations and actions, and compress the temporal evolution of the incident into a single snapshot. 
In this example, the reachability fault precedes the observed impact and is both topologically and functionally distant: the impact occurs on data-plane routers, while the root cause originates in control-plane components (controllers), separated by two hops through a hierarchical aggregation node.

This example illustrates the difficulty of identifying true root causes in large-scale cloud networks, where causally related events may be separated in time, space, and abstraction level, and where simple proximity- or correlation-based reasoning can be misleading.

In this paper, we propose \textbf{\method}, a learning-based system for root cause attribution based on counterfactual reasoning \footnote{Throughout the paper, the term ``counterfactual'' is used in the wide sense of reasoning about the impact of hypothetical interventions -- in contrast to the ``ladder of causation'' in \cite{Pearl2018}, which strictly distinguishes between interventional (rung 2) and counterfactual (rung 3) causal reasoning.}. Our key insight is to train a generative model that captures the forward dynamics of fault propagation in
network systems, and then leverage this model as a simulator for interventional analysis. Specifically, we learn the conditional distribution $P(X_{t+1} | X_t, G)$ where $X_t$ encodes node-level fault and action states and $G$ represents the incident subgraph. At inference time, we perform counterfactual simulation by removing candidate faults from historical sequences and rolling the model forward to predict the resulting impact. The divergence between factual and interventional outcomes provides a principled measure of causal influence.

This work is motivated by four key objectives for practical RCA in large-scale networks.
First, it must be \emph{principled}, providing causal attribution grounded in counterfactual reasoning rather than correlation alone.
Second, it must be \emph{adaptive}, learning spatiotemporal fault propagation patterns directly from data instead of relying on hand-crafted rules that are brittle under evolving network conditions.
Third, it must be \emph{effective} in realistic operational settings, improving early root cause ranking where only a small number of hypotheses can be acted upon.
Finally, it must be \emph{feasible at scale}, both computationally and operationally: the approach should operate efficiently on incident-specific subgraphs and avoid reliance on large, expert-labeled training datasets, which are prohibitively expensive to obtain in production network environments.

\method  meets these objectives by design, and improves exact-match root cause identification by 16.1\% over a rule-based heuristic on expert-labeled production incidents. Despite the complexity of counterfactual evaluation, inference remains practical: a sequential CPU-only implementation achieves sub-minute response times at the 95th percentile, and the approach readily supports parallelization and GPU batching for substantially lower latency.

Our contributions are as follows:
\begin{itemize}
\item \formulation: We formulate network root cause analysis as a causal inference problem over graph-temporal systems, enabling principled causal attribution via interventional simulation.
\item \model: We develop a generative spatiotemporal model that captures fault propagation dynamics in heterogeneous network graphs without requiring labeled root causes for training.
\item \metric: We introduce the Total Causal Influence (TCI) metric, which ranks candidate root causes by measuring divergence between factual and interventional outcomes.
\item \effectiveness: We evaluate our approach on large-scale production data from a leading cloud provider, including an expert-labeled benchmark of real incidents.
\end{itemize}

\section{Related Work}

We position \method at the intersection of (i) causal inference and counterfactual reasoning on graph-temporal data, (ii) graph machine learning for learning spatiotemporal dynamics, and (iii) root cause analysis in large-scale networked systems. Collectively, existing approaches address subsets of these challenges, but no single method simultaneously provides all of the capabilities required for practical, incident-level causal attribution in large-scale networks. Table~\ref{tab:rca-comparison} provides a concise summary of these trade-offs.

\subsection{Causal inference and counterfactual reasoning on graphs and time series}

Classical causal inference represents causal structure via directed acyclic graphs and Structural Causal Models (SCMs), where interventions enable counterfactual reasoning and causal attribution~\cite{pearl2009causality}. Inferring complete causal graphs and structural equations from data, however, requires strong assumptions and is particularly challenging in highly complex environments such as cloud networks~\cite{montagna2023assumption}. \method does not aim to recover a global SCM; instead, it learns a forward transition mechanism that can be queried under counterfactual perturbations, yielding a pragmatic notion of causal influence aligned with operational RCA.

Several approaches learn graph-structured dynamics from data. Neural Relational Inference (NRI)~\cite{kipf2018neural} jointly infers latent interactions and temporal dynamics, but is primarily optimized for forecasting and structure discovery rather than intervention-based attribution. DECI~\cite{bica2021deci} estimates causal effects in longitudinal data via counterfactual outcome modeling, focusing on treatment effect estimation over fixed variables. In contrast, \method assumes a known heterogeneous topology and applies counterfactual simulation to rank candidate causes within individual incidents, rather than estimating population-level causal effects.

\subsection{Graph machine learning for spatiotemporal dynamics and world models}

Graph neural networks (GNNs) are widely used to model dynamics in interacting systems, where message passing captures how local interactions give rise to global behavior~\cite{gilmer2017neural,li2018diffusion,prakash2012spotting}. More recent work explores spatiotemporal GNNs and graph-based simulators that learn transition dynamics conditioned on graph structure, enabling multi-step prediction and reasoning over time~\cite{zhao2025tgnns}.

Relatedly, \emph{world models} learn latent dynamics that support prediction and counterfactual reasoning without explicit physical state representations~\cite{ha2018worldmodels}. \method adopts a similar philosophy in the context of network operations, learning a spatiotemporal transition model over sparse fault and action signals grounded in known network topology. Unlike prior work focused on forecasting, control, or planning, \method uses the learned simulator explicitly for \emph{incident-level counterfactual analysis}, evaluating how predicted impact changes when specific historical fault transitions are removed.

\subsection{Root cause analysis in networked and distributed systems}

RCA has long been studied in network and systems management. Early approaches relied on rule-based correlation and codebooks~\cite{kliger1995coding,yemini1996high}, which are effective for known failure modes but require substantial expert effort and struggle to generalize. More recent data-driven methods apply causal graphs and machine learning to metrics and logs, particularly in microservice systems~\cite{xin2022causalrca,root_cause_analysis,Hardt2023petshop,zheng2024mulan}. These approaches demonstrate the value of causal modeling, but they typically operate over service or metric graphs (using metric data from the non-anomalous regime \cite{orchard2025root}) and focus on dependency discovery and localization rather than explicit simulation of incident dynamics.

Graph-based learning has also been applied to RCA as supervised node ranking or classification over KPI graphs~\cite{yen2022graph}, which requires labeled data at scale and stable graph structure. Other work combines GNNs with Granger-style causality or fault-tree assumptions in controlled environments~\cite{liu2023granger_gat}, limiting applicability to open-ended production networks. Large language models have recently been explored as RCA assistants~\cite{chen2024rcacopilot,roy2024llmrca}, but primarily support investigation workflows with causal hypotheses that are not transparent~\cite{liu2024llmnet_survey}.

\method differs from prior RCA approaches in three key respects: it models incidents as heterogeneous graph-temporal structures grounded in physical, logical, and hierarchical topology; it learns fault propagation dynamics directly from incident data without requiring large labeled RCA datasets; and it performs attribution via counterfactual simulation, providing an explicit, model-based measure of causal influence.

\paragraph*{Positioning summary}
Table~\ref{tab:rca-comparison} summarizes how representative RCA approaches trade off causal grounding, scalability, heterogeneity, and operational actionability. By combining learned spatiotemporal simulation with counterfactual evaluation over real network topology, \method bridges classical causal reasoning and modern graph learning to enable practical, incident-level root cause attribution in large-scale networks.

\begin{table}[t]
\centering
\small
\resizebox{0.85\linewidth}{!}{
\begin{tabular}{l|ccccc||c}
\toprule
\textbf{Capability} &
\rotatebox{90}{Rule-based~\cite{kliger1995coding}} &
\rotatebox{90}{Correlation~\cite{yemini1996high}} &
\rotatebox{90}{Causal Graph~\cite{xin2022causalrca, root_cause_analysis, Hardt2023petshop, zheng2024mulan}} &
\rotatebox{90}{Supervised GNN~\cite{bica2021deci, yen2022graph}} &
\rotatebox{90}{LLM-based~\cite{chen2024rcacopilot, roy2024llmrca, liu2024llmnet_survey}} &
\rotatebox{90}{\method} \\
\midrule
Explicit counterfactual intervention      &  &  & \mycheck &  &  & \mycheck \\
Temporal dynamics                         &  & \mycheck & \mycheck & \mycheck &  & \mycheck \\
Topology-aware modeling                  & \mycheck & \mycheck & \mycheck & \mycheck &  & \mycheck \\
Heterogeneous entities/relations         &  &  & \mycheck &  &  & \mycheck \\
No large labeled RCA data needed          & \mycheck & \mycheck & \mycheck &  & \mycheck & \mycheck \\
Low/no upfront expert effort needed       &  & \mycheck &  & \mycheck  &  & \mycheck \\
Incident-level attribution               & \mycheck &  & \mycheck & \mycheck & \mycheck & \mycheck \\
Direct actionability                     & \mycheck &  &  &  & \mycheck & \mycheck \\
\bottomrule
\end{tabular}
}
\caption{\shortTag{\method positioning:} Comparison of RCA approaches across key operational capabilities.}
\label{tab:rca-comparison}
\end{table}

\section{Problem Formulation}

We formalize root cause analysis as a \emph{causal attribution} problem over graph-temporal incident data. Given a customer impact detected by observability systems (e.g., packet loss or elevated latency), our objective is to identify which faults in the incident \emph{causally explain} the observed impact, enabling targeted mitigation and remediation. We assume access to (i) observability systems that detect impacts and report fault and action signals, and (ii) comprehensive topology data capturing network entities, their physical and logical interconnections, and hierarchical relationships, as is typical in well-monitored cloud networks.

More concretely, given an incident represented as a network graph $G$, with a customer impact observed at time $T$ and a sequence of time-indexed fault transitions and actions preceding the impact, the problem is to identify and rank those transitions in the incident history that \emph{causally explain} the observed impact.

\subsection{Incident Graph Construction and Representation}\label{sec:incident-graph}

Each incident is initialized from a known impact: a set of impact nodes and a timestamp at which data-plane degradation was observed. We construct a seeded incident subgraph by recursively incorporating spatial neighbors observed within a temporal window prior to impact, following Layer-3 links, optical connections, and hierarchical containment edges. This yields a variable-sized, heterogeneous graph that captures the incident-relevant portion of the system. In large cloud networks, unconstrained topological expansion can introduce many nodes that are unlikely to be causally relevant. To address this, we allow configuration of which device types are included in the incident graph based on the impact modality and its location. For each node in the resulting graph, we extract impact, impairment, and action signals from our observability and automation systems.

Formally, we represent each incident as an undirected heterogeneous graph
\[
G = (V, E, \tau, \phi),
\]
where $V$ is the set of nodes, $E \subseteq V \times V$ is the set of edges,
$\tau : V \rightarrow \mathcal{Q}$ maps each node to a node type (e.g., device, site, metro),
and $\phi : E \rightarrow \mathcal{R}$ maps each edge to a relationship type
(e.g., physical, logical, hierarchical).

Nodes in $V$ represent network entities at multiple levels of abstraction (e.g., devices, sites/data centers, metros), while edge types encode interaction and containment relationships such as physical connectivity, logical adjacencies, and hierarchy (device--site, site--metro).

The graph structure $G$ is static for the duration of an incident.
Temporal evolution is captured through node-level feature matrices
$\{X_t\}_{t=0}^{T}$, where each $X_t$ is a random matrix taking values in
$\{0,1\}^{|V| \times F}$ and represents the latent fault and action state at time $t$
(typically at minute-level resolution). We denote by $x_t$ a realized observation
of $X_t$ for a specific incident. In our implementation, $F = 27$, comprising 15 fault dimensions (e.g., device hardware error, link failure, optical degradation) and 12 action dimensions (e.g., traffic drain, maintenance activity, configuration change).

\subsection{State Transitions and Causal Attribution}\label{sec:state-transitions}

We define a \textbf{state transition} as any change in a feature from 0 to 1 at time $t$:
\[
(v, f, t) \ \ \text{where}\ \ x_t[v,f] = 1 \ \text{and}\ x_{t-1}[v,f] = 0.
\]

Let $S = \{(v, f, t)\}$ denote the set of all observed state transitions during an incident. We partition this set into:
\begin{itemize}
\item \textbf{Impact set} $I \subseteq S$: Transitions occurring at the final impact time $T$, representing customer-visible degradation.
\item \textbf{Hypothesis set} $H \subseteq S \setminus I$: Prior transitions potentially responsible for the impact, i.e., $H = \{(v, f, t) \in S \mid t \leq T\}$.
\end{itemize}

For a subset of incidents, we assume access to an expert-labeled ground-truth root cause set $r \subseteq H$, representing the faults that led to the observed impact. The objective is to predict a candidate set $\hat{r} \subseteq H$ that explains the impact.

We define two correctness criteria:
\begin{itemize}
\item \textbf{Weak correctness}: $\hat{r} \cap r \neq \emptyset$ (at least one true root cause identified).
\item \textbf{Strong correctness}: $\hat{r} = r$ (all qualified root causes identified).
\end{itemize}
In our evaluation, we focus on weak correctness, as identifying at least one true root cause is often sufficient to trigger the correct mitigation or remediation action in operational settings.

\section{Methodology}

\method consists of (a) a world model (a generative spatiotemporal model that predicts future node states conditioned on incident structure and history), and (b) a counterfactual simulation procedure that ranks candidate root-cause hypotheses via their induced effect on predicted impact. This design enables scalable learning without reliance on large root-cause-labeled datasets.

\subsection{Generative Spatiotemporal Model}\label{sec:generative-model}

The goal of the generative model is to predict the next-step state of each node from current states, conditioned on the incident graph:
\[
\hat{x}_{t+1} = f_\theta(x_{0:t}, G),
\]
the output $\hat{x}_{t+1}$ denotes per-node logits (or probabilities) for the next-step binary state. The architecture comprises four components (Figure~\ref{fig:netcause-nn-arch}):

\begin{figure}[t]
\centering
\resizebox{0.8\linewidth}{!}{
\begin{tikzpicture}[
    font=\small,
    node distance=9mm and 12mm,
    box/.style={draw, rounded corners, align=center, minimum width=27mm, minimum height=8mm},
    sbox/.style={draw, rounded corners, align=center, minimum width=32mm, minimum height=9mm},
    arrow/.style={-{Latex[length=2.2mm,width=1.6mm]}, line width=0.6pt},
    dashedarrow/.style={-{Latex[length=2.2mm,width=1.6mm]}, line width=0.6pt, dashed}
]

\node[box] (x) {$X_t$\\\footnotesize (binary node states)};
\node[box, right=of x] (g) {$G$\\\footnotesize (incident graph)};

\node[sbox, below=of x, xshift=18mm] (emb) {Feature\\Embeddings\\\footnotesize (fault/action types)};

\node[sbox, below=of emb] (spat) {Spatial Encoder\\(R-GCN $\times L$)\\\footnotesize hetero message passing};

\node[sbox, below=of spat] (temp) {Temporal Encoder\\(RNN / GRU)\\\footnotesize per-node history};

\node[sbox, below=of temp] (dec) {Decoder\\\footnotesize logits for $\hat{X}_{t+1}$};

\node[box, below=of dec] (y) {$\hat{X}_{t+1}$\\\footnotesize (predicted next state)};

\draw[arrow] (x) -- (emb);
\draw[arrow] (emb) -- (spat);
\draw[arrow] (g) |- (spat);

\draw[dashedarrow] (emb.west) -- ++(-10mm,0) |- (temp.west);

\draw[arrow] (spat) -- (temp);

\draw[arrow] (temp) -- (dec);
\draw[arrow] (dec) -- (y);

\node[font=\footnotesize, align=left, left=6mm of temp, yshift=-2mm
] (fusion) {Fusion: \\ spatial embedding \\ + feature skip};

\end{tikzpicture}
}

\caption{\textbf{\method generative spatiotemporal model architecture.} At each timestep $t$, the model consumes binary node states $X_t$ and the incident graph $G$. Feature embeddings provide a learned representation of fault/action types. A relational GCN performs heterogeneous message passing over $G$ to produce spatial embeddings, which are combined (via a skip connection) with embedded features and processed by a temporal encoder to model per-node dynamics. A decoder outputs logits for the next state $\hat{X}_{t+1}$.}
\label{fig:netcause-nn-arch}
\end{figure}
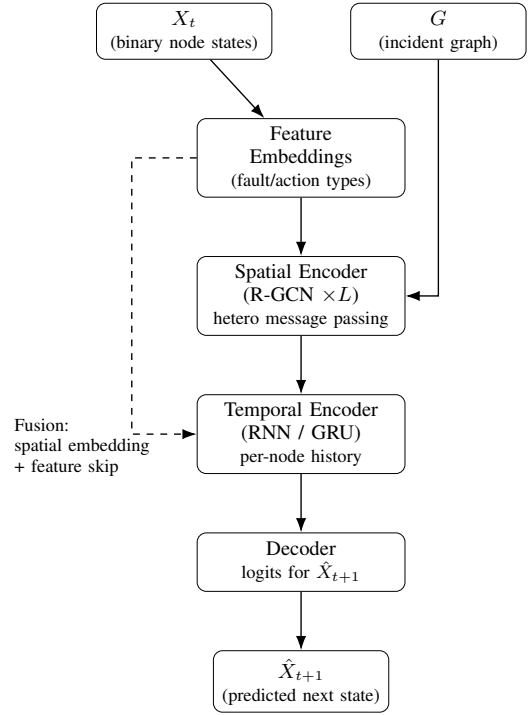

\textbf{Feature Embeddings.} We introduce learnable embeddings for each fault and action type. Each of the $F$ features is associated with a trainable vector, which is combined with the raw binary features to provide a more expressive input representation.

\textbf{Spatial Encoder.} We use a Relational Graph Convolutional Network (R-GCN) to propagate information across the heterogeneous incident graph. R-GCN enables the model to learn distinct propagation patterns for physical, logical, and hierarchical relationships. We stack multiple spatial layers to allow information flow across multiple hops.
The final spatial layer produces feature-wise state transitions (using an activation function)  to be used in subsequent temporal encoding.

\textbf{Temporal Encoder.} We apply a recurrent neural network (RNN) independently to each node's history (with shared parameters across nodes) to capture temporal dependencies. We also use a skip connection that injects the embedded node features directly into the temporal encoder alongside the spatial embedding, improving gradient flow and model stability.

\textbf{Decoder.} The decoder produces per-node logits for the binary features of $x_{t+1}$, and the model is trained using binary cross-entropy with positive-class weighting to address severe sparsity.

\textbf{Training Procedure.} We train with teacher forcing, feeding the observed state $x_t$ as input to predict
$\hat{x}_{t+1}$ (instead of using the previous prediction), and apply multi-step unrolling to capture longer-range propagation.

\textbf{Loss Function.} We optimize the model using binary cross-entropy with logits, applied to the predicted and observed next-step states. To address the severe class imbalance inherent in sparse fault and action signals, we apply positive-class weighting, which increases the penalty for false negatives relative to false positives.

\subsection{Counterfactual Simulation and Causal Attribution}

At inference time, the trained world model serves as a simulator of incident evolution under alternative historical assumptions. For each candidate hypothesis $\eta = (v, f, t_\eta) \in H$, we construct a counterfactual sequence in which all observed events are held except the hypothesized fault transition\footnote{Here we assume that there are no hidden common causes that influence both present and past states of the system. This assumption of {\it unconfoundedness} of the dynamics is required to conclude that the {\it interventional} probability of  $X_{t+1}$, given hypothetical changes of $X_t$ coincides with the {\it observational} probability, given $X_t$.}:
\[
x^{\mathrm{cf}} = x \setminus \{x_{v,f,t_\eta} = 1\}.
\]

We roll the model forward from $t_\eta$ to the impact time $T$, producing predicted interventional states $\{\hat{x}^{\mathrm{cf}}_t\}_{t=t_\eta}^{T}$. Hypotheses that causally contribute to the incident are expected to induce larger deviations between factual and counterfactual predictions.

We define the \textbf{Total Causal Influence} (TCI) as the cumulative divergence between factual and counterfactual predictions over the remaining evolution of the incident:
\[
\mathrm{TCI}(\eta)
= \sum_{t=t_\eta}^{T}\ \sum_{v' \in V}\ w(t)\,\mathcal{D}\!\left(\hat{x}_t,\ \hat{x}^{\mathrm{cf}}_t\right),
\]
where $\mathcal{D}(\cdot,\cdot)$ is a divergence measure (in our implementation, we use the absolute difference in first-order Markov log-likelihood assigned to the two realizations, though alternatives such as total variation distance or F1-based divergence are possible), and $w(t)$ is a temporal weighting function allowing us to emphasize effects at different time scales (we chose an exponential weighting function to more heavily weight signals closer to the impact time).
\[
\mathcal{D}\!\left(\hat{x}_t,\hat{x}^{\mathrm{cf}}_t\right)
=
\left|
\log p_\theta(x_t \mid x_{0:t-1},G)
-
\log p_\theta(x_t \mid x^{\mathrm{cf}}_{0:t-1},G)
\right|.
\]
Candidates with TCI below a configurable threshold are discarded as unlikely root causes, and the remaining hypotheses are ranked by TCI to produce a prioritized list. Computing TCI requires evaluating counterfactual rollouts for each hypothesis. For an incident with $|H|$ hypotheses, $T$ timesteps, and $|V|$ nodes, the computational complexity is $\mathcal{O}(|H|\cdot T\cdot |V|)$. In practice, this is tractable because (i) the generative model produces predictions for all nodes/features in a single forward pass per timestep, (ii) incidents are sparse, and (iii) counterfactual simulations can be batched and parallelized on GPUs.


\section{Experimental Setup}

\subsection{Training Data}

To train the generative spatiotemporal model, we extract 1,500 operational incidents from the production network of a leading cloud provider. Incidents are seeded from detected packet-loss impacts across multiple metropolitan regions over a six-month period, covering a diverse set of network topologies, traffic patterns, and failure modes.

For each detected impact, we construct an incident-specific subgraph using the procedure described in Section~\ref{sec:incident-graph}, incorporating physical, logical, and hierarchical relationships among network entities. The resulting incident graphs vary in size and structure, reflecting the localized scope of individual incidents. Model training follows the procedure described in Section~\ref{sec:generative-model}. Additional details on hyperparameter selection and training stability are provided in the Appendix.

\subsection{Evaluation Benchmark}

To evaluate root cause attribution performance, we construct a benchmark of 31 incident graphs labeled with ground-truth root causes based on post-incident troubleshooting reports and expert assessment by network engineers. These labels identify the faults that were determined to have directly caused the observed customer impact.

On average, each labeled incident contains 5.4 ground-truth root causes and 67.25 observed fault and action transitions that constitute candidate hypotheses. This corresponds to an average root cause proportion of 9.5\%, indicating that fewer than one in ten observed signals is causally responsible for the impact. This imbalance reflects realistic operational conditions, where many signals are present but only a small subset is relevant for diagnosis.

Table~\ref{tab:root-cause-breakdown} summarizes the distribution of labeled incidents by root cause category. While most incidents are associated with a single category, some involve multiple valid root causes (e.g., a control-plane error accompanied by a hardware packet error). Link-related issues are the most prevalent category, accounting for over one-third of incidents, which is expected given the scale of optical infrastructure and routine maintenance activities in large cloud networks.

\begin{table}[t]
\centering
\small
\resizebox{0.5\linewidth}{!}{
\begin{tabular}{lc}
\toprule
\textbf{Root Cause Category} & \textbf{Count} \\
\midrule
LinkIssue & 13 \\
Congestion & 6 \\
HardwarePacketError & 5 \\
StateUpdate & 3 \\
ReachabilityError & 2 \\
ProtocolError & 1 \\
Rebuild & 1 \\
ReturnToService & 1 \\
ControlPlaneError & 1 \\
\bottomrule
\end{tabular}
}
\caption{Distribution of root cause categories across 31 labeled incidents.
\label{tab:root-cause-breakdown}}
\end{table}

\subsection{Baseline Methods}

We compare \method against three baseline methods that reflect common operational heuristics used in network troubleshooting. All baselines operate on the same incident graphs and candidate hypothesis sets.

\textbf{Temporal Proximity.} Hypotheses are ranked by temporal proximity to the impact, prioritizing faults that occurred closest in time to the observed customer degradation. This baseline captures the common intuition that recent events are more likely to be causal.

\textbf{Spatial Proximity.} Hypotheses are ranked by graph distance to impact nodes, computed using weighted shortest paths over the incident graph. Edge weights reflect relationship types (physical, logical, hierarchical). This baseline captures the intuition that faults topologically closer to the impact location are more likely to be responsible.

\textbf{Rule-Based Heuristic.} We implement a composite scoring function inspired by operator practice, combining temporal, spatial, and event-type signals:
\[
\text{score} = 0.1 \cdot d_{\text{norm}} + 0.6 \cdot t_{\text{norm}} + 0.3 \cdot s_{\text{event}},
\]
where $d_{\text{norm}}$ is normalized graph distance, $t_{\text{norm}}$ is normalized time difference from impact (clipped at one hour), and $s_{\text{event}}$ is a severity score assigned based on fault type using operator-provided guidelines. Lower scores indicate higher likelihood of being a root cause.

\subsection{Evaluation Metrics}\label{sec:eval}

Each method produces a ranked list of candidate hypotheses for each incident. We evaluate performance using the following metrics, where $k$ denotes the number of top-ranked hypotheses considered. We cap $k$ at 5, as larger candidate sets provide diminishing practical value for operators and automation systems.

\textbf{Exact Match Accuracy.} The fraction of incidents for which the top-ranked hypothesis matches at least one ground-truth root cause, following the \textit{Weak Correctness} criterion defined in Section~\ref{sec:state-transitions}. This metric reflects automation readiness, where only the highest-confidence diagnosis is acted upon.

\textbf{Hits@k.} The fraction of incidents for which at least one ground-truth root cause appears among the top-$k$ ranked hypotheses. This metric captures whether a method is able to surface a correct explanation within a small candidate set, regardless of ordering.

\textbf{Precision@k.} The fraction of hypotheses in the top-$k$ rankings that correspond to ground-truth root causes, averaged across incidents. This measures the purity of the ranked list.

\textbf{Recall@k.} The fraction of ground-truth root causes recovered within the top-$k$ ranked hypotheses, averaged across incidents. This measures coverage of relevant causes.

For recall and precision, we normalize by $\min(k, |r_i|)$ for each incident $i$, where $|r_i|$ is the number of ground-truth root causes, ensuring fair comparison across incidents with varying label counts. Only exact root cause matches are considered correct; correlated but downstream signals are not counted. For the last three metrics, we also compute the confidence interval at one standard deviation using a nonparametric bootstrap method with 1,000 samples.

\section{Results}

We evaluate \method on 31 expert-labeled production incidents and compare against the three baseline methods discussed earlier. Figures~\ref{fig:exact-match}--\ref{fig:recall-at-k} summarize the results.

\subsection{Exact Match Accuracy}

\begin{figure}[t]
\centering
\includegraphics[width=0.9\columnwidth]{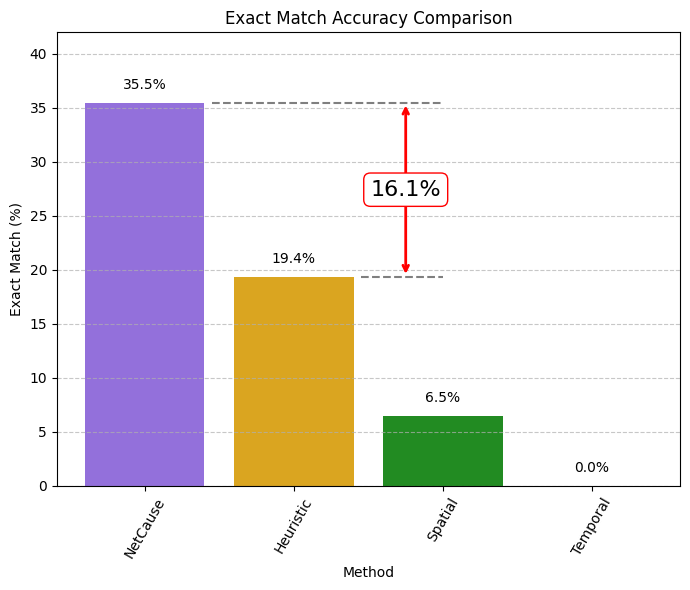}
\caption{\shortTag{\method wins at exact match accuracy}. We achieve 35.5\% accuracy in identifying a correct root cause as the top-ranked hypothesis.}
\label{fig:exact-match}
\end{figure}

Figure~\ref{fig:exact-match} shows that \method achieves an exact match accuracy of 35.5\% (11 out of 31 incidents), representing a 16.1 percentage point absolute improvement and an 82\% relative improvement over the rule-based heuristic. This result highlights improved automation readiness: in more than one-third of incidents, the top-ranked hypothesis produced by \method directly identifies a true root cause. In contrast, the spatial and temporal proximity baselines perform poorly on this metric, achieving 6.5\% and 0.0\% accuracy, respectively. This confirms that neither temporal recency nor topological proximity alone is sufficient for reliable root cause attribution.

\subsection{Hits@k}

\begin{figure}[t]
\centering
\includegraphics[width=0.9\columnwidth]{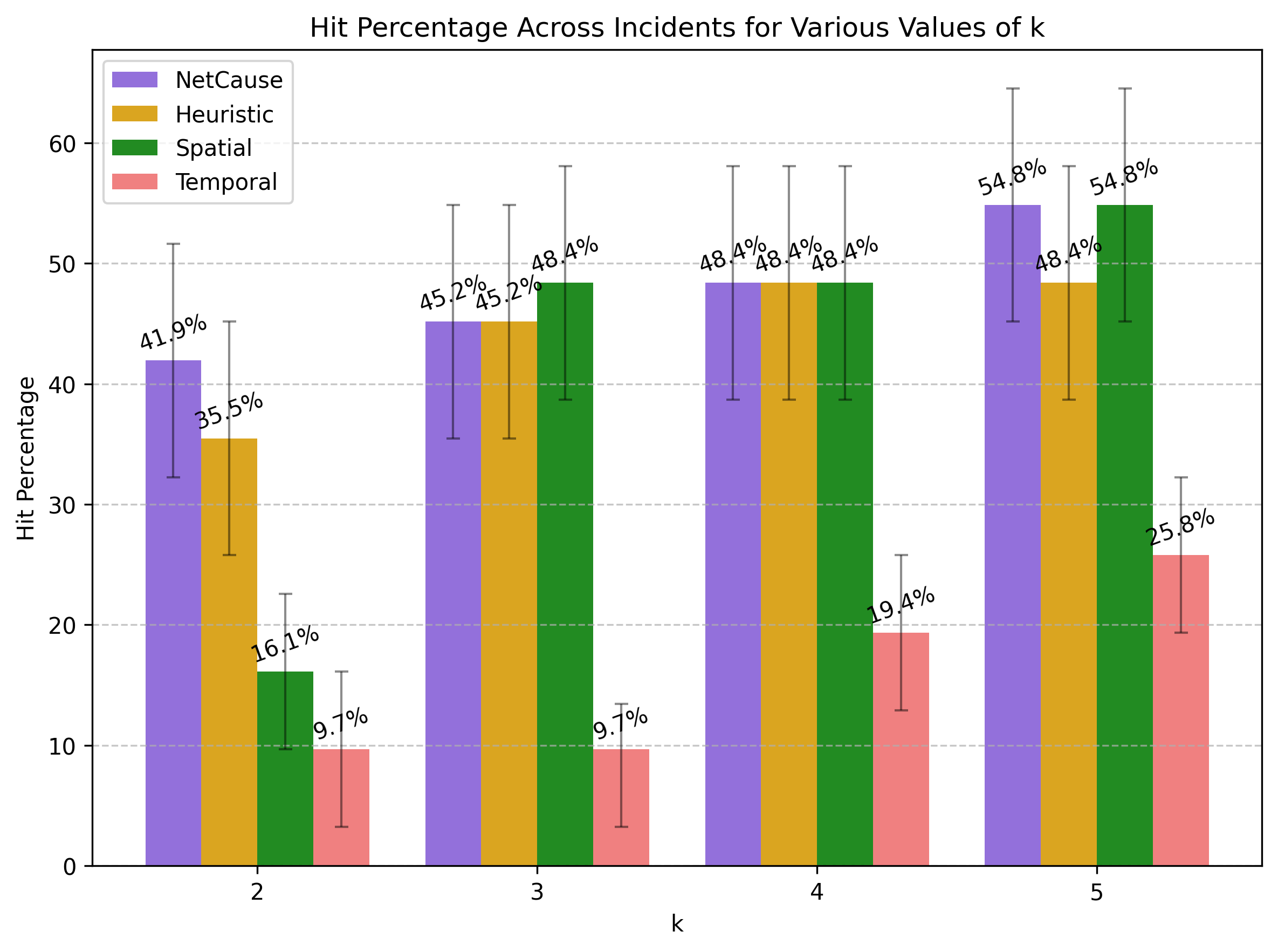}
\caption{\shortTag{\method wins} almost always: Hits@k showing the fraction of incidents with at least one correct root cause in the top-$k$ ranked hypotheses. \method in purple.}
\label{fig:hit-at-k}
\end{figure}

Figure~\ref{fig:hit-at-k} reports Hits@k, measuring whether at least one correct root cause appears among the top-$k$ hypotheses for each incident. As expected, performance improves with increasing $k$ for all methods. At $k=2$, \method achieves Hits@2 of 41.9\%, outperforming the rule-based heuristic (35.5\%). At $k=3$ and $k=4$, \method and the heuristic method exhibit similar performance and are joined by the spatial baseline, indicating that spatial proximity becomes increasingly informative when larger candidate sets are considered. At $k=5$, both \method and the spatial baseline successfully surface a correct root cause in 17 incidents (54.8\%), while the rule-based heuristic does so in 15 incidents (48.4\%). The temporal baseline reaches only 25.8\%.

\subsection{Precision@k}

\begin{figure}[t]
\centering
\includegraphics[width=0.9\columnwidth]{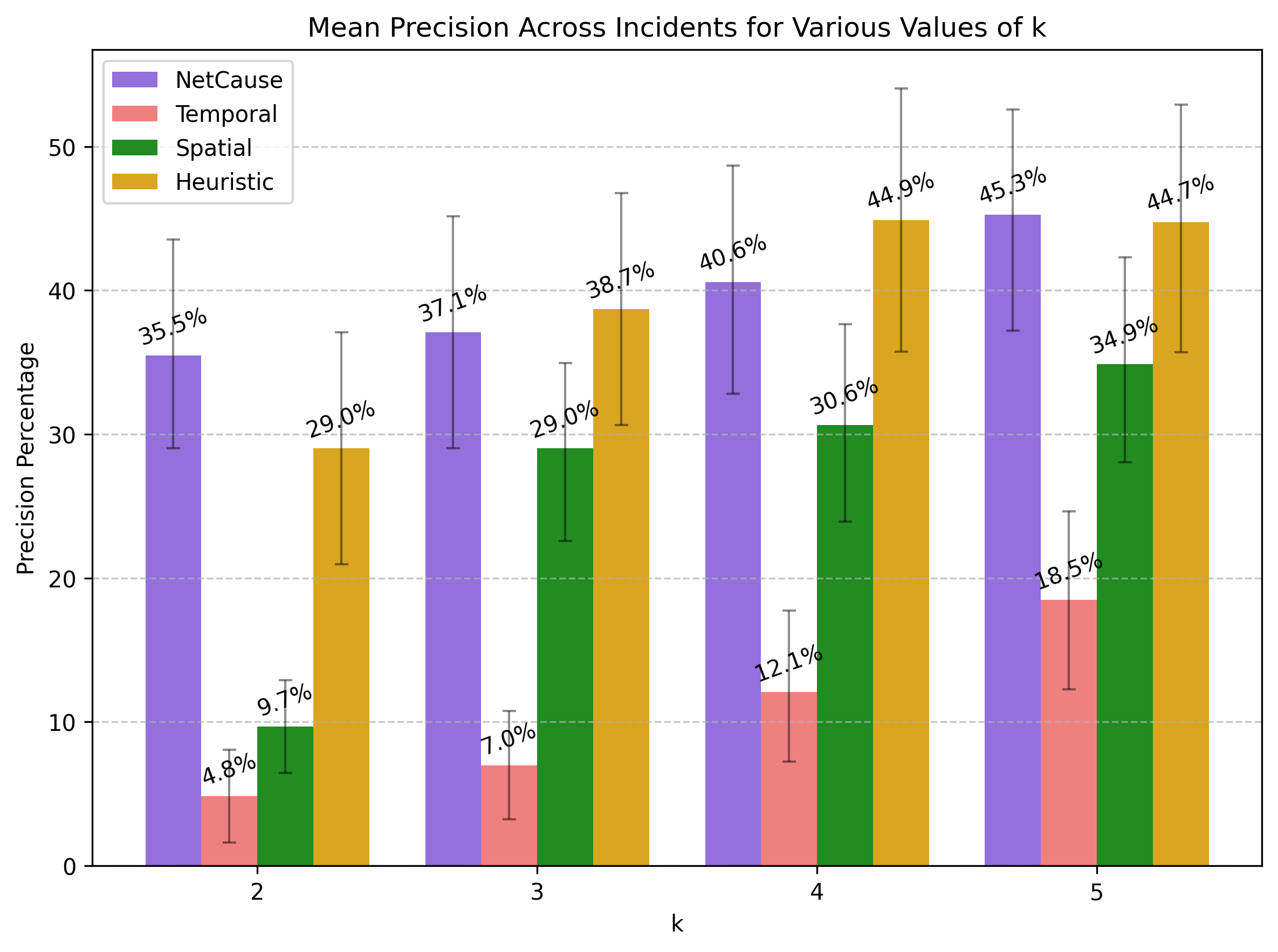}
\caption{Precision@k showing the fraction of top-$k$ hypotheses that correspond to ground-truth root causes, averaged across incidents. \method in purple.}
\label{fig:precision-at-k}
\end{figure}

Figure~\ref{fig:precision-at-k} shows that \method achieves higher precision at small values of $k$ compared to the other methods. At $k=2$, \method attains a precision of 35.5\%, indicating that approximately one-third of the hypotheses presented to operators are correct on average. For larger values of $k$, \method and the rule-based heuristic exhibit similar precision. At $k=5$, both methods achieve precision of approximately 45\%. This trend reflects a trade-off between recall and precision: as $k$ increases, more correct hypotheses are surfaced, but at the cost of introducing additional non-causal signals.

\subsection{Recall@k}

\begin{figure}[t]
\centering
\includegraphics[width=0.9\columnwidth]{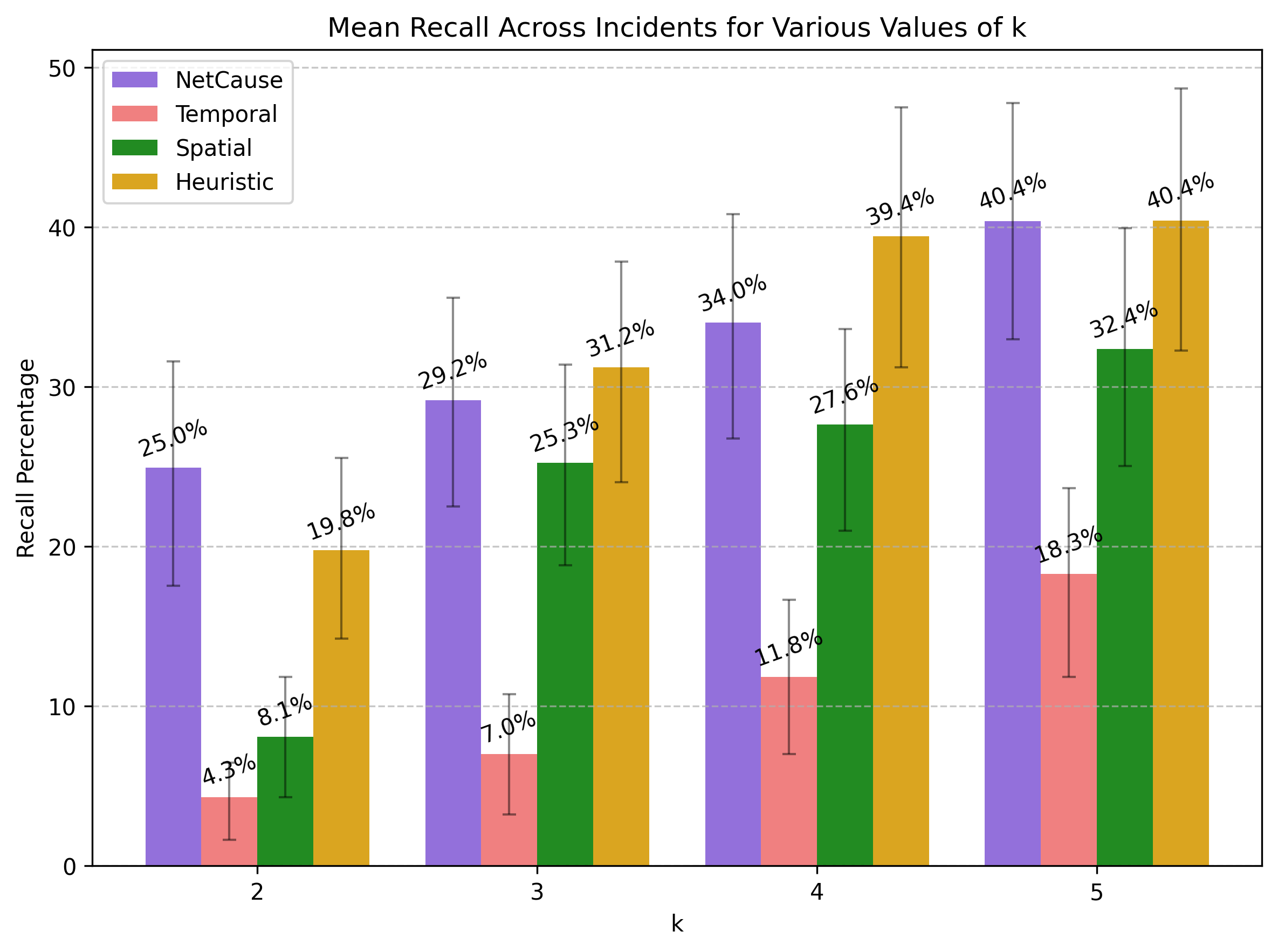}
\caption{Recall@k showing the fraction of ground-truth root causes recovered within the top-$k$ ranked hypotheses. \method in purple.}
\label{fig:recall-at-k}
\end{figure}

Figure~\ref{fig:recall-at-k} presents Recall@k. \method outperforms baseline methods at $k=2$, recovering approximately 25\% of all ground-truth root causes, compared to 20\% for the rule-based heuristic. At larger values of $k$, \method and the heuristic method converge, each recovering approximately 40\% of root causes by $k=5$. The spatial and temporal baselines underperform across all values of $k$.

\section{Discussion}

The results highlight a consistent pattern across evaluation metrics: \method provides its strongest gains at small values of $k$, particularly for top-ranked predictions. In terms of exact match accuracy and performance at $k \in \{1,2\}$, \method substantially outperforms temporal, spatial, and rule-based baselines. This indicates that counterfactual reasoning over learned graph-temporal dynamics improves the quality of hypothesis ranking when only a small number of diagnostic actions can be taken. In operational settings, this regime is especially important, as automation systems and on-call engineers typically act on one or two high-confidence hypotheses rather than exploring long candidate lists.

At the same time, \method does not dominate all baselines across all values of $k$. For larger candidate sets ($k \geq 3$), the rule-based heuristic and spatial proximity baseline achieve comparable performance. This convergence is expected: as more hypotheses are considered, simple heuristics based on temporal and topological proximity increasingly capture at least one causally relevant signal. These baselines effectively exploit strong priors that are broadly valid in network incidents, even if they lack principled causal grounding.

Several factors likely contribute to the observed performance gap between \method and the baselines not widening further. First, both the training corpus (1{,}500 incidents) and the labeled evaluation benchmark (31 incidents) are relatively small by the standards of deep learning. In this regime, model performance is sensitive to data composition, and removing or adding a small number of incidents can noticeably affect aggregate metrics. We expect that scaling the training set to several thousand additional incidents would improve the robustness and ranking accuracy of the learned generative model.

Second, the input data itself is noisy and incomplete. The fault and action signals used by \method are not raw representations of network state, but outputs of observability systems designed to manage alert volume through thresholds, suppression rules, and timing constraints. As a result, these signals may be delayed, suppressed, or only indirectly related to the true underlying cause. In some cases, the true root cause may not be explicitly observable at all. During labeling, we selected the most upstream signal available in the causal chain, but this may still reflect a proxy rather than the originating failure. These limitations affect all evaluated methods and constrain the achievable performance ceiling.

Despite these factors, the results indicate that \method offers meaningful advantages over rule-based and heuristic approaches in realistic operational settings. While heuristic methods remain effective for many simple and recurring incidents, they become increasingly brittle as incident complexity grows and failure modes evolve. In contrast, \method learns fault propagation patterns directly from data and evaluates candidate causes through counterfactual simulation, reducing reliance on manual rule specification and improving robustness in complex, multi-fault scenarios.

Importantly, \method complements rather than replaces existing RCA techniques. It improves early root cause ranking and provides principled causal attribution, while still benefiting from simple temporal and topological priors when broader exploration is possible. This makes it a practical component for production network operations and automated remediation workflows.

\method is also computationally feasible for deployment. Inference achieves sub-30 second latency at the 95th percentile using a sequential CPU-only implementation, which is well within operational constraints given minute-level telemetry. Because counterfactual simulations are independent, inference can be parallelized via batching or GPU execution to further reduce latency (sub-second).

\paragraph*{Limitations and sensitivity}Our approach relies on several assumptions and design choices that may affect performance in practice. First, evaluation is conducted on a relatively small set of 31 expert-labeled incidents, which introduces statistical variability in aggregate metrics; we partially address this by reporting confidence intervals, but larger labeled benchmarks would provide stronger validation. Second, the counterfactual attribution mechanism assumes approximate unconfoundedness of the observed dynamics; in real deployments, unobserved factors or delayed signals may violate this assumption and affect attribution accuracy. Third, the effectiveness of the TCI metric depends on the fidelity of the learned generative model and the choice of temporal weighting function; while we use an exponential weighting to emphasize signals near the impact, alternative choices may yield different rankings. Empirically, we observe that performance is stable for moderate weighting parameters ($[0.1, 1.0]$), with degradation only at larger values, suggesting that the method is not overly sensitive to the exact choice of temporal weighting within a reasonable range. Finally, our formulation focuses on $0 \rightarrow 1$ state transitions, which simplifies modeling but does not explicitly capture recovery dynamics; while challenging from a telemetry point-of-view, extending the model to also reason over fault resolution ($1 \rightarrow 0$) is a promising direction for future work.

\section{Conclusion}

We presented \method, a learning-based framework for root cause analysis in large-scale networked systems that combines graph-temporal modeling with counterfactual simulation. Our method is designed to meet four key requirements for practical RCA in modern cloud networks, as we discussed earlier: \emph{Principled}, \emph{Adaptive}, \emph{Effective} and \emph{Feasible at scale}. Our evaluation on expert-labeled production incidents shows that \method improves the quality of root cause ranking in the regime most relevant to automation and operational decision-making. When only a small number of hypotheses can be acted upon, \method achieves stronger early ranking and more reliable top-1 and top-2 predictions than heuristic baselines. At larger candidate set sizes, performance converges with heuristic methods, reflecting the continued usefulness of simple temporal and topological priors when broader exploration is possible.

Beyond empirical gains, \method offers a complementary path forward for network operations. Rule-based and heuristic approaches remain effective for many simple and recurring incidents but become increasingly brittle as system complexity grows. By learning propagation dynamics and validating candidate causes through counterfactual evaluation, \method reduces manual maintenance burden while preserving interpretability and actionability. Taken together, these results suggest that counterfactual reasoning over learned graph-temporal models is a promising foundation for scalable, data-driven, and actionable root cause analysis in large-scale network management and automation.

\bibliographystyle{IEEEtran}
\bibliography{stcr_icccn}

\appendix
\subsection{Model Details: R-GCN Spatial Encoder}
\label{app:rgcn}

For completeness, we provide the Relational Graph Convolutional Network (R-GCN) update used in the spatial encoder.
Let $h_v^{(l)} \in \mathbb{R}^{d}$ denote the hidden representation of node $v$ at layer $l$, and let $\mathcal{N}_m(v)$
denote the set of neighbors of $v$ under edge type $m \in M$. A standard R-GCN layer computes:
\[
h_v^{(l+1)} = \sigma\!\left(
\sum_{m \in M} \sum_{u \in \mathcal{N}_m(v)} \frac{1}{c_{v,m}}\, W_m^{(l)} h_u^{(l)} \;+\; W_0^{(l)} h_v^{(l)}
\right),
\]
where $W_m^{(l)} \in \mathbb{R}^{d \times d}$ are relation-specific weight matrices, $W_0^{(l)} \in \mathbb{R}^{d \times d}$
is a self-loop weight matrix, $\sigma(\cdot)$ is a non-linear activation, and $c_{v,m}$ is a normalization constant (e.g.,
$c_{v,m} = |\mathcal{N}_m(v)|$). Stacking $L$ such layers enables multi-hop propagation across the heterogeneous incident graph.

\subsection{Model Architecture and Hyperparameter Optimization}

We conducted extensive hyperparameter optimization to identify architectures that generalize well to unseen incidents. Our search space included:
\begin{itemize}
\item Number of R-GCN layers (hops): $\{1, 2, 3, 4\}$
\item Hidden dimension: $\{8, 16, 32, 64, 128, 256, 512\}$
\item Number of RNN layers: $\{1, 2, 4, 8\}$
\item RNN type: $\{\text{RNN}, \text{LSTM}, \text{GRU}\}$
\item Positive class weight: $\{0.5, 1.0, 1.5, 2.0, 2.5, 10, 50, 100\}$
\end{itemize}

Key findings from our optimization:
\begin{enumerate}
\item \textbf{Compact models generalize better.} Smaller hidden dimensions (8-16) and fewer RNN layers (1) consistently outperformed larger architectures. This suggests that the fault propagation patterns in our data are relatively low-dimensional and that overparameterization leads to overfitting.

\item \textbf{Low positive class weights are optimal.} Weights of 0.5-2.0 achieved the best F1 scores, indicating that aggressive upweighting of rare events degrades performance. This likely reflects the fact that not all positive examples are equally informative.

\item \textbf{Moderate spatial depth is sufficient.} While 4-hop R-GCN models showed better overall performance and generalization, 2-hop models achieved higher peak F1 scores on some incidents. We selected 4 hops as the default to balance generalization and peak performance.

\item \textbf{Simple RNN outperforms LSTM/GRU.} Vanilla RNN cells performed comparably or better than gated variants, suggesting that the temporal dependencies in our data do not require complex gating mechanisms.

\item \textbf{Bidirectional graphs are critical.} Adding reverse edges for hierarchical relationships improved F1 scores substantially (from ~0.5 to ~0.87).
\end{enumerate}

Our final architecture uses: 4 R-GCN layers, hidden dimension 16, 1 RNN layer, vanilla RNN cells, positive weight 0.5, bidirectional graph schema, feature embeddings with Xavier initialization, and skip connections.

\subsection{Time Bucketization of Incident Graph}
Node features are represented as sparse binary time series, recording only state transitions to avoid redundant time steps. This results in irregular temporal sampling where consecutive observations may be separated by varying intervals. To handle this, we discretize time into non-uniform buckets centered around the impact time, with finer resolution near the impact. This reflects the empirical observation that temporal proximity is a strong cue for fault propagation while reducing data dimensionality and model size. The choices of node and edge types, and time bucketization capture some of our assumptions and biases that we impose on the model.


\end{document}